\documentclass[english]{extarticle}
\usepackage[T1]{fontenc}
\usepackage[latin9]{inputenc}
\usepackage{amsthm}
\usepackage{amsmath}

\makeatletter
\numberwithin{equation}{section}
\numberwithin{figure}{section}
\newcommand{\lyxaddress}[1]{
\par {\raggedright #1
\vspace{1.4em}
\noindent\par}
}

\makeatother

\usepackage{babel}

\begin{document}

\title{Entanglement Entropy for the Charged BTZ Black Hole}

\author{Alexis Larrañaga}

\maketitle

\lyxaddress{Observatorio Astronomico Nacional. Universidad Nacional de Colombia}
\begin{abstract}
Using the AdS/CFT correspondence we calculate the explicit form of
the entanglement entropy for the charged BTZ black hole. The leading
term in the large temperature expansion of the entropy function for
this black hole reproduces its Bekenstein-Hawking entropy and the
subleading term, representing the first corrections due to quantum
entanglement, behaves as a logarithm of the BH entropy. 
\end{abstract}

\section{Introduction}

Recently, the idea of holography has gained a great attention. It
claims that the degrees of freedom in $\left(d+2\right)$-dimensional
quantum gravity will be comparable to those of quantum many body systems
in $\left(d+1\right)$ dimensions. This was essentially found by remembering
that the entropy of a black hole is not proportional to its volume,
but to the area its event horizon (Bekenstein-Hawking entropy),

\begin{equation}
S^{BH}=\frac{A}{4G},\end{equation}
where $G$ is Newton constant. 

The discovery of the AdS/CFT correspondence show explicit examples
where the holography is manifestly realized. This correspondence states
that the quantum gravity on $(d+2)$-dimensional anti-de Sitter spacetime
($\mbox{AdS}_{d+2}$) is equivalent to a certain conformal field theory
in $d+1$ dimensions ($\mbox{CFT}_{d+1}$). However, the fundamental
mechanism of the AdS/CFT correspondence is not known today.

In this paper we investigate quantum entanglement in the context of
three-dimensional (3D) AdS gravity, in particular of the charged Bañados-Teitelboim-Zanelli
(BTZ) black hole\cite{chargedBTZ}, using the AdS3/CFT2 correspondence.
In order to obtain the entropy function we will use the standard method
for studying correlations in QFT described in \cite{cadoni} in which
two external length-scales are introduced in the boundary 2D CFT.
One is a thermal wavelength $\beta=\frac{1}{T}$ (with $T$ the CFT's
temperature) and a spatial length $\gamma$ which is the measure of
the observable spatial region of the 2D universe.

\section{Entanglement Entropy}

Consider a $2D$ spacetime with a compact spacelike dimension with
$S^{1}$ topology and length $\ell$. If the only part of the universe
accesible for measurement is a spacelike slice $\Sigma$ with length
$\gamma$, we loose information about the degrees of freedom localized
in the complementary region $\Sigma'$. The entanglement entropy originated
by tracing over the unobservable degrees of freedom is given by the
von Neumann entropy,

\begin{equation}
S^{ent}=-\mbox{Tr}_{\Sigma}\left[\rho_{\Sigma}\ln\rho_{\Sigma}\right],\end{equation}
where $\rho_{\Sigma}$ is the reduced density matrix, obtained by
tracing the density matrix $\rho$ over the states of the $\Sigma'$
region,

\begin{equation}
\rho_{\Sigma}=\mbox{Tr}_{\Sigma'}\rho.\end{equation}

The entanglement entropy for the ground state of the $2D$ CFT at
zero temperature, with a spacelike dimension with $S^{1}$ topology
and infinite timelike direction (cylinder), is given by \cite{cadoni,reviewEE}

\begin{equation}
S_{cyl}^{ent}=\frac{c+\bar{c}}{6}\ln\left(\frac{\ell}{\epsilon\pi}\sin\frac{\pi\gamma}{\ell}\right),\end{equation}
where $c$ and $\bar{c}$ are the central charges of the $2D$ CFT
and $\epsilon$ is an ultra-violet cutoff used to regularize the divergence
that appears because of the sharp boundary between the regions $\Sigma$
and $\Sigma'$.

When $\ell\gg\gamma$, the spacelike dimension becomes infinite and
the entanglement entropy is independient of $\ell$. Therefore, the
entanglement entropy for a $2D$ CFT at zero temperature on a plane
is given by

\begin{equation}
S_{cyl}^{ent}=\frac{c+\bar{c}}{6}\ln\left(\frac{\gamma}{\epsilon}\right).\label{eq:zerotemp}\end{equation}

However, if we consider a $2D$ CFT at finite temperature $T=\frac{1}{\beta}$
on a plane, the entanglement entropy is

\begin{equation}
S_{cyl}^{ent}=\frac{c+\bar{c}}{6}\ln\left(\frac{\beta}{\epsilon\pi}\sinh\frac{\pi\gamma}{\beta}\right).\end{equation}

As is well known, from the correspondence between $3D$ AdS gravity
and $2D$ CFT ($\mbox{AdS}_{3}/\mbox{CFT}_{2}$) there follows that
the entanglement entropy of a $2D$ CFT contains information about
quantum gravity correlations of its $3D$ gravity holographical dual.
In this particular case, the CFT has a central charge given by \cite{brown,reviewEE}

\begin{equation}
c=\bar{c}=\frac{3L}{2G},\label{eq:c}\end{equation}
where $L$ is the AdS length and $G$ is the $3D$ newtonian constant.

\section{The Charged BTZ Black Hole}

The (2+1)-dimensional BTZ (Banados-Teitelboim-Zanelli) black holes
have obtained a great importance in recent years because the provide
a simplified model for exploring some conceptual issues, not only
about black hole thermodynamics \cite{btz,btz1} but also about quantum
gravity and string theory. The charged BTZ black hole is a solution
of the (2+1)-dimensional gravity theory with a negative cosmological
constant $\Lambda=-1/L^{2}$ . The metric is given by\cite{chargedBTZ}
\begin{equation}
ds^{2}=-f(r)dt^{2}+\frac{dr^{2}}{f(r)}+r^{2}d\varphi^{2},\label{eq:metric}\end{equation}
 where \begin{equation}
f(r)=-M+\frac{r^{2}}{L^{2}}-\frac{Q^{2}}{2}\ln\left[\frac{r}{L}\right]\end{equation}
 is known as the lapse function and $M$ and $Q$ are the mass and
electric charge of the BTZ black hole, respectively. The lapse function
vanishes at the radii $r=r_{\pm}$, where $r_{+}$ gives the position
of the event horizon. The electric potential of this black hole is 

\begin{equation}
\Phi=\frac{\partial M}{\partial Q}=-Q\ln\left[\frac{r}{L}\right]\label{eq:potential}\end{equation}

while the Hawking temperature is given, as usual, by \begin{equation}
T_{H}=\frac{1}{4\pi}\left|\frac{df(r)}{dr}\right|_{r=r_{+}}=\frac{r_{+}}{2\pi L^{2}}-\frac{Q^{2}}{8\pi r_{+}},\end{equation}
and its entropy by

\begin{equation}
S_{BTZ}^{BH}=\frac{A}{4G}=\frac{\pi r_{+}}{2G}.\end{equation}

The spinless charged BTZ black hole can be considered as the thermalization
at temperature $T=T_{H}=\frac{1}{\beta_{H}}$ of the charged AdS spacetime.
On the 2D boundary of the AdS spacetime, and in the large temperature
limit $\left(r_{+}\gg L\right)$, this thermalization corresponds
to a plane/cylinder transformation that maps the CFT on the plane
in the CFT on the cylinder. The conformal map plane/cylinder has the
form

\begin{equation}
z=\exp\left[\frac{2\pi}{\beta_{H}}\left(\varphi+it\right)\right].\end{equation}
The above transformation is the asymptotic form of the map between
the BTZ black hole and $\mbox{AdS}_{3}$ in Poincaré coordinates,
and corresponds to the conformal transformation that maps the entanglement
entropy of a CFT at zero temperature (on the plane) in that of a CFT
at finite temperature (in the cylinder). Correspondingly, the holographic
entanglement entropy of the charged AdS spacetime becomes the holographic
entanglement entropy of the BTZ black hole 

\begin{eqnarray}
\tilde{S}_{BTZ}^{ent} & = & S_{CFT}^{ent}\left(\gamma=2\pi L,\beta=\beta_{H}\right)\\
 & = & \frac{c}{3}\ln\left[\frac{8r_{+}L^{2}}{\epsilon\left(4r_{+}^{2}-L^{2}Q^{2}\right)}\sinh\left(\frac{\pi r_{+}}{L}-\frac{\pi LQ^{2}}{4r_{+}}\right)\right].\end{eqnarray}

This entanglement entropy depends on the UV cutoff given by the parameter
$\epsilon$. To obtain a renormalized entropy we will subtract the
contribution of the vacuum (i.e. the zero mass, zero temperature BTZ
black hole solution). This is given by equation (\ref{eq:zerotemp}),

\begin{equation}
S_{vac}^{ent}=S_{cyl}^{ent}\left(\gamma=2\pi L\right)=\frac{c}{3}\ln\left(\frac{2\pi L}{\epsilon}\right).\end{equation}

Therefore, the renormalized entanglement entropy is given by

\begin{equation}
S_{BTZ}^{ent}=\tilde{S}_{BTZ}^{ent}-S_{vac}^{ent}\end{equation}

\begin{equation}
S_{BTZ}^{ent}=\frac{c}{3}\ln\left[\frac{4r_{+}L}{\pi\left(4r_{+}^{2}-L^{2}Q^{2}\right)}\sinh\left(\frac{\pi r_{+}}{L}-\frac{\pi LQ^{2}}{4r_{+}}\right)\right],\end{equation}
or using (\ref{eq:c}),

\begin{equation}
S_{BTZ}^{ent}=\frac{L}{2G}\ln\left[\frac{4r_{+}L}{\pi\left(4r_{+}^{2}-L^{2}Q^{2}\right)}\sinh\left(\frac{\pi\left(4r_{+}^{2}-L^{2}Q^{2}\right)}{4r_{+}L}\right)\right].\label{eq:entropy}\end{equation}

Note that, as expected, the renormalized entanglement entropy goes
to zero as $4r_{+}^{2}-L^{2}Q^{2}\rightarrow0$ or $r_{+}\rightarrow\frac{LQ}{2}$
(this is the condition for the charged BTZ black hole ground state).
The entanglement entropy for the charged BTZ black hole reduces to
the entropy obtained for the spinless BTZ solution reported in \cite{cadoni}
and using the relation between 2D and 3D Newton constant,

\begin{equation}
G_{2}=\frac{L}{4G},\end{equation}
it reproduces exactly the result of \cite{cadoni2}. 

In the large temperature limit $\left(r_{+}\gg L\right)$, equation
(\ref{eq:entropy}) can be expanded as 

\begin{equation}
S_{BTZ}^{ent}\approx\frac{\pi r_{+}}{2G}-\frac{\pi}{2G}\frac{L^{2}Q^{2}}{4r_{+}}-\frac{L}{2G}\ln\left[\frac{\pi\left(4r_{+}^{2}-L^{2}Q^{2}\right)}{4r_{+}L}\right]+O\left(\frac{L}{r_{+}}\right)^{2}\label{eq:aux1}\end{equation}

\begin{equation}
S_{BTZ}^{ent}\approx\frac{\pi r_{+}}{2G}-\frac{L}{2G}\ln\left[\frac{\pi r_{+}}{L}\right]+O\left(\frac{L}{r_{+}}\right)\end{equation}

\begin{equation}
S_{BTZ}^{ent}\approx S_{BTZ}^{BH}-\frac{L}{2G}\ln\left[\frac{\pi r_{+}}{L}\right]+O\left(\frac{L}{r_{+}}\right).\end{equation}

The first term in the expansion is the Bekenstein-Hawking entropy
for the BTZ black hole, $S_{BTZ}^{BH}=\frac{\pi r_{+}}{2G}$, and
describes the situation in which thermal fluctuations dominates, i.e.
the entanglement entropy becomes the thermodynamical entropy in this
limit. The second term describes the first corrections to the entropy
due to quantum entanglement. This behavior reproduces results of Cadoni
and Melis\cite{cadoni}, as well as the logarithmic term obtained
in \cite{kumar,mann}. 

It is also interesting that the second term in (\ref{eq:aux1}) is
an inverse of area term in subleading order similar to the one obtained
by S. K . Modak in the equation (63) of\cite{kumar} for the rotating
BTZ black hole and also in the general case studied by Akbar and Saifullah
\cite{key-1}. In this paper, the authors calculate the entropy of
the black hole considering the quantum tunneling of paricles through
the event horizon, using two unknown parameters, $\alpha_{1}$ and
$\alpha_{2}$. The present analysis let us fix those parameters for
the charged BTZ black hole as

\begin{eqnarray}
\alpha_{1} & = & \frac{L}{8\pi G}\\
\alpha_{2} & = & \frac{L^{3}Q^{2}}{64G}.\end{eqnarray}

\section{Conclusion}

We have derived the entanglement entropy function for the charged
BTZ black hole, obtaining an expansion with a leading term that corresponds
to the thermodynamical entropy and subleading terms that describe
the corrections due to quantum entanglement. This result sheds light
on the meaning of the holographic entanglement entropy for the charged
BTZ black hole. The resulting function for the entropy seems to have
a meaning only in the regime $r_{+}\gg L$, i.e. for macroscopic black
holes.

\emph{Acknowledgements}. This work was supported by the Universidad
Nacional de Colombia. Project Code 2010100.

\end{document}